\documentclass[11pt]{article}

\usepackage{sectsty}
\usepackage{graphicx}
\usepackage{subcaption}
\usepackage{authblk}
\title{Revolutionizing MRI Data Processing Using FSL: Preliminary Findings with the Fugaku Supercomputer}
\author[1]{Tianxiang Lyu}
\author[1,2]{Wataru Uchida}
\author[1,2]{Zhe Sun}
\author[1,2]{Christina Andica}
\author[1,2]{Keita Tokuda}
\author[1]{Rui Zou}
\author[1]{Jie Mao}
\author[1,3]{Keigo Shimoji}
\author[1,3]{Koji Kamagata}
\author[1,2]{Mitsuhisa Sato}
\author[1,2]{Ryutaro Himeno}
\author[1,2,3]{Shigeki Aoki}
\affil[1]{Graduate School of Medicine, Juntendo University}
\affil[2]{Faculty of Health Data Science, Juntendo University}
\affil[3]{Department of Radiology, Juntendo University}
\date{\today}

\begin{document}
\maketitle

\section{Abstract}
The amount of Magnetic resonance imaging data has grown tremendously recently, creating an urgent need to accelerate data processing, which requires substantial computational resources and time. In this preliminary study, we applied FMRIB Software Library commands on T1-weighted and diffusion-weighted images of a single young adult using the Fugaku supercomputer. The tensor-based measurements and subcortical structure segmentations performed on Fugaku supercomputer were highly consistent with those from conventional systems, demonstrating its reliability and significantly reduced processing time.
\section{Introduction}
Magnetic resonance imaging (MRI) has revolutionized neuroscience research, enabling comprehensive analyses of brain structure, function, and connectivity in both healthy individuals and those with neurological or psychiatric conditions. To advance our understanding of the brain and its disorders, large-scale neuroimaging datasets like the Human Connectome Project (HCP)\cite{glasser_human_2016}, the UK Biobank (UKB)\cite{littlejohns_uk_2020}, the Adolescent Brain Cognitive Development (ABCD)\cite{cetin-karayumak_harmonized_2024}, and Brain/MINDS-beyond have been established. However, the sheer volume of data in these datasets poses a significant computational challenge, hindering research progress and limiting the scope of analysis.\cite{koike_brainminds_2021}
\par
Traditional methods for processing brain MRI data often rely on resources available at the university laboratory level, which are insufficient to handle the massive scale of these datasets. The processing time required for thousands of brain images becomes impractical, leading to delays in research projects and potentially limiting the insights that can be derived from these valuable datasets. Supercomputers offer a promising solution by enabling high-speed, parallel processing of large datasets\cite{jimbo_accelerated_2023}, but deploying complex MRI data processing pipelines on these systems remains a challenge due to issues such as software dependencies, hardware-software compatibility, and the complexity of composite pipelines involving multiple software packages.
\par
Previous attempts to implement MRI data processing pipelines on supercomputers have been limited in scope. While studies have successfully implemented the FreeSurfer pipeline\cite{jimbo_accelerated_2023}, which accepts T1-weighted images as input, on the supercomputer Fugaku, there are no known studies that have implemented the FMRIB Software Library (FSL) pipeline, which can accept both T1-weighted and diffusion-weighted images as input. The ability to process diverse types of MRI data on a supercomputer would dramatically increase the options for brain image analysis and accelerate research progress.
\par
To run FSL on Fugaku supercomputer, The main difficulty is complication. FSL official do not provide a correct configuration setting for Fugaku supercomputer. The reason should be obivous that FSL offcial have no chance accessing Fugaku before, and all codes in FSL have been compiled and run for many years on Intel-based platform only, therefore it is hard to launch on Fugaku supercomputer. Therefore, our main efforts is to modify the source codes and configure files of FSL, in order to make it able to be compiled and run on Fugaku supercomputer. The main challenges are shown in the following:
\begin{itemize}
\item Configuration files were written 10 years ago, where the system could not guess build type.
\item The software dependencies on Fugaku are not compatible with FSL.
\item Some source codes are written with Intel-based CPU SIMD instrunctions only.
\end{itemize}
\par
This study aims to address these challenges by demonstrating the feasibility of utilizing the FSL pipeline on the Fugaku supercomputer system. By showcasing the successful integration of FSL onto a leading-edge supercomputer, we seek to pave the way for efficient and scalable analysis of large-scale neuroimaging datasets. The unique capabilities of the Fugaku supercomputer, such as its high-performance computing power and ability to handle complex workflows, make it particularly suitable for this research. The successful implementation of the FSL pipeline on Fugaku has the potential to revolutionize brain research by enabling researchers to process and analyze vast amounts of diverse MRI data efficiently, ultimately leading to new insights into brain structure, function, and disorders.
\section{Method}
\subsection{Study Participant}
The MRI data of a 26-year-old female subject analyzed in this study was randomly selected from the S1200 dataset of the WU-Minn Human Connectome Project (HCP) consortium,\cite{VanEssen2013} which consists of high-quality neuroimaging data acquired from a large cohort of healthy young adults. Upon further examination, the subject was found to have normal cognitive function as indicated by a Mini-Mental State Examination score of 30, no vascular risk factors (normal body mass index, blood pressure, lipid, and glucose profiles), and no history of neurological or psychological diseases, alcohol abuse, or drug abuse.
\subsection{Data Acquisition}
Imaging data from the HCP database were acquired using an HCP-customized Siemens 3T Connectome Skyra magnet.\cite{VanEssen2012} Diffusion-weighted images were acquired using a spin-echo echo-planar imaging sequence with the following parameters: repetition time (TR) = 5520 $ms$, echo time (TE) = 89.5 $ms$, flip angle = 78°, field of view (FOV) = 210 × 180 $mm$, matrix size = 144 × 124, and voxel size = 1.5 × 1.5 × 1.5 $mm^3$. Diffusion weighting was applied along 90 diffusion-encoding directions at b = 1000, 2000, and 3000 $s/mm^2$, with an additional six b = 0 $s/mm^2$ (non-diffusion-weighted) images acquired. From the available diffusion-weighted data, only the volumes with b = 0 and 1000 s/mm$^2$ were extracted and utilized for analysis in this study, because the current work focused exclusively on diffusion tensor imaging (DTI) analyses. At higher b-values (2000 and 3000 s/mm$^2$), the diffusion signal becomes increasingly influenced by non-Gaussian motion effects, violating the Gaussian diffusion assumption underlying the DTI model. T1-weighted images were acquired using a 3D Magnetization-Prepared Rapid Gradient-Echo sequence with the following parameters: TR = 2400 $ms$, TE = 2.14 $ms$, flip angle = 8°, FOV = 224 × 224 $mm$, matrix size = 320 × 320, slice thickness = 0.7 $mm$, and voxel size = 0.7 × 0.7 × 0.7 $mm^3$. All imaging data were processed and reconstructed using the HCP's minimal preprocessing pipelines,\cite{Glasser2013} which included gradient nonlinearity correction, motion correction, and registration to the MNI anatomical template. On visual examination, T1 and diffusion-weighted images of the subject assessed in this study were free from artifacts and structural abnormalities.
\subsection{Processing Pipelines}
The analysis pipeline was implemented using tools from FSL version 6.0,\cite{Jenkinson2021} which provides a comprehensive library of brain image analysis techniques including segmentation, registration, and mathematical operations. The specific steps shown in Figure 1 were as follows:
\begin{enumerate}
\item Estimation of DTI parameters from diffusion-weighted images using DTIFIT, including Fractional Anisotropy (FA) [unitless], Mean Diffusivity (MD) [$\times 10^{-3}$ mm$^2$/s], Axial Diffusivity (AD) [$\times 10^{-3}$ mm$^2$/s], and Radial Diffusivity (RD) [$\times 10^{-3}$ mm$^2$/s].
\item Segmentation of subcortical structures from T1-weighted images using FMRIB's Integrated Registration and Segmentation Tool (FIRST).\cite{Patenaude2011} Subcortical structures to be evaluated included accumbens, amygdala, brainstem, caudate, hippocampus, pallidum, putamen, and thalamus.
\item Linear registration of the T1-weighted image to diffusion space using FMRIB's Linear Image Registration Tool (FLIRT) (Jenkinson and Smith, 2001) and epi\_reg. The masks of subcortical regions obtained from FIRST were then transformed to diffusion space using nearest neighbor interpolation.
\item Computation of average DTI parameter values within each subcortical region using fslstats.
\end{enumerate}
\begin{figure}[h!] % image examples & compare
	\centering
    \includegraphics[width=0.8\textwidth]{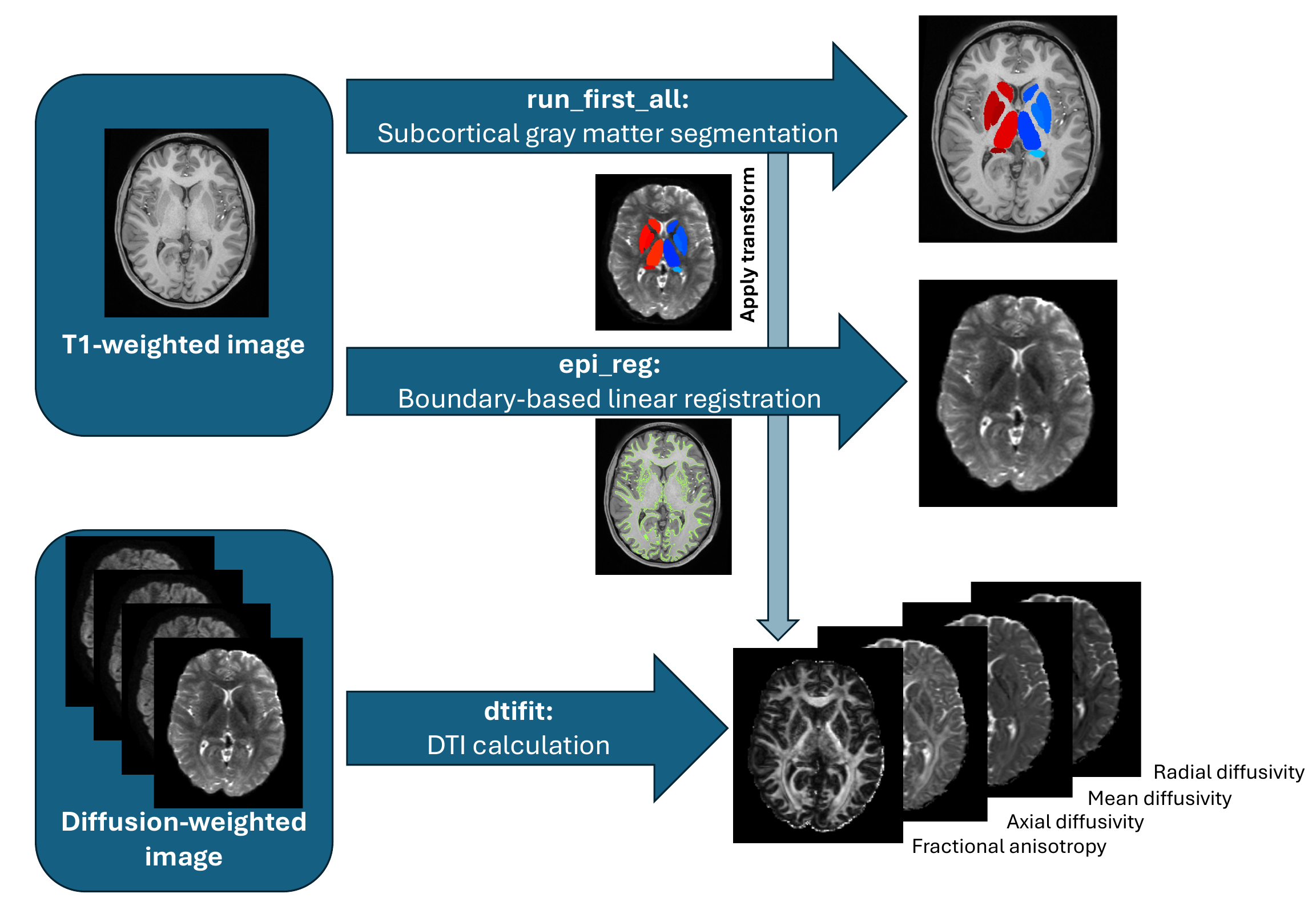}
    \caption{Processing pipeline set on supercomputers}
    \label{fig:brain_atlas}
\end{figure}
\subsection{System Configuration}
The environment used in this study is solely FMRIB Software Library (FSL) version 6.0.5, and the platforms are shown as the following. Because it is the 

The analysis results of version 7.2.0 were not examined by Haddad et al.21; thus, we present new test–retest results.
\subsubsection{Fugaku Supercomputer}
The supercomputer Fugaku is a homogeneous, CPU-based distributed machine with 158,976 nodes. Based on the ARM v8.2A architecture, the CPU of the system, A64FX, has 32 GB HBM2 memory and 48 compute cores, with total memory bandwidth 1024 GB/s. 12 compute cores form one Core Memory Group (CMG). On the ALUs side, there is a couple of 512-bit wide SIMD units inside each compute core, where the whole CPU performs 3.072 TF/s for double-precision, 6.144 TF/s for single-precision, and 12.288 TF/s for half-precision. As to cache, Each compute core contains 64 KB of L1 data cache, and 32 MB L2 cache is shared by all. For interconnect, low latency and high throughput are achieved by a 6-dimensional mesh/torus network, with a 6.8 GB/s link bandwidth and 40.8 GB/s of injection bandwidth per compute node.
\begin{itemize}
\item Operating System: Red Hat Enterprise Linux 8 McKernel
\item Arch: Armv8.2-A SVE 512 bit
\item Chip: Fujitsu A64FX
\item CPU: 48 compute cores and 2-4 assistant cores with HBM2 32 GiB, 1024 GB/s
\item Memory: 4.85 PiB HBM2
\end{itemize}
To run FSL on Fugaku supercomputer, The main difficulty is handle the compatibility of software environment. However, FSL official do not provide a correct configuration setting for Fugaku supercomputer. Therefore, it is critical to build up an environment with usable configuration. Our current software environment is shown as the following:
\begin{itemize}
\item GNU Compiler: 8.5.0 
\item OpenBLAS: 0.3.21
\item Lapack: 3.10.8
\end{itemize}
\subsubsection{Conventional node}
To evaluate the reproducibility of our brain image analysis software implemented on a supercomputer, we used a MacBook Pro as the comparison node. The specifications of the MacBook Pro are as follows:
\begin{itemize}
\item Operating System: macOS 14.2
\item Model: MacBook Pro (Model Identifier: MacBookPro18,4)
\item Chip: Apple M1 Max
\item CPU: 10-core (8 performance cores and 2 efficiency cores)
\item Memory: 64 GB LPDDR5
\end{itemize}
These specifications provided a baseline for comparing the performance and reproducibility of the analysis software when run on a high-performance supercomputer versus a conventional node. Because there is official configuration of FSL on macOS, the software environment is managed by FSL install script.
\subsection{Data Parallelism Strategies}
In this section, we will talk about the strategies of data parallel processing on supercomputers, which is designed for processing large numbers of brain image with high efficiency, considering not only the CPU architecture but also the available computing resource of the whole system.
\subsubsection{NUMA-Aware Optimizations}
HBM memory offer higher memory bandwidth, but there are several NUMAs in one CPU. Data transfer inside one NUMA can achieve highest memory bandwidth with lowest response latency. Therefore, if memory size is available, one processing pipelines should be performed in one NUMA in order to achieve best performance. Also, most of processing libraries are not supported with multi-threading parallelization, which makes it much more important to execute several pipelines on one CPU, in order to utilize multiple computing cores.\\
Inside one computing node, GNU Parallel library is used to execute pipelines in parallelism. And the maximal subjects processed on one compute node is related to the memory consumption of executed pipelines.\\
\subsubsection{Bulk Jobs}
There are hundreds to thousands of computing node in supercomputers, offering the promise of expressive speedups over single workstation. One of the main technical contribution of our paper is to utilize massive computing nodes using bulk jobs for scheduling. Bulk jobs decomposes a whole data processing into many subtasks, wherer there is no dependency or communication required between the execution of the subtasks. Then all subtask will be submitted to the waiting list of the workload manager. where each subtask will be executed with available computing resource. Such an parallel strategy can be readily deployed on leading-edge supercomputers and experimental evidence shows that such parallelization introduces more flexibility and efficiency of the computing resource on leading-edge supercomputers.
\begin{figure}[h!] % image examples & compare
	\centering
    \includegraphics[width=0.9\textwidth]{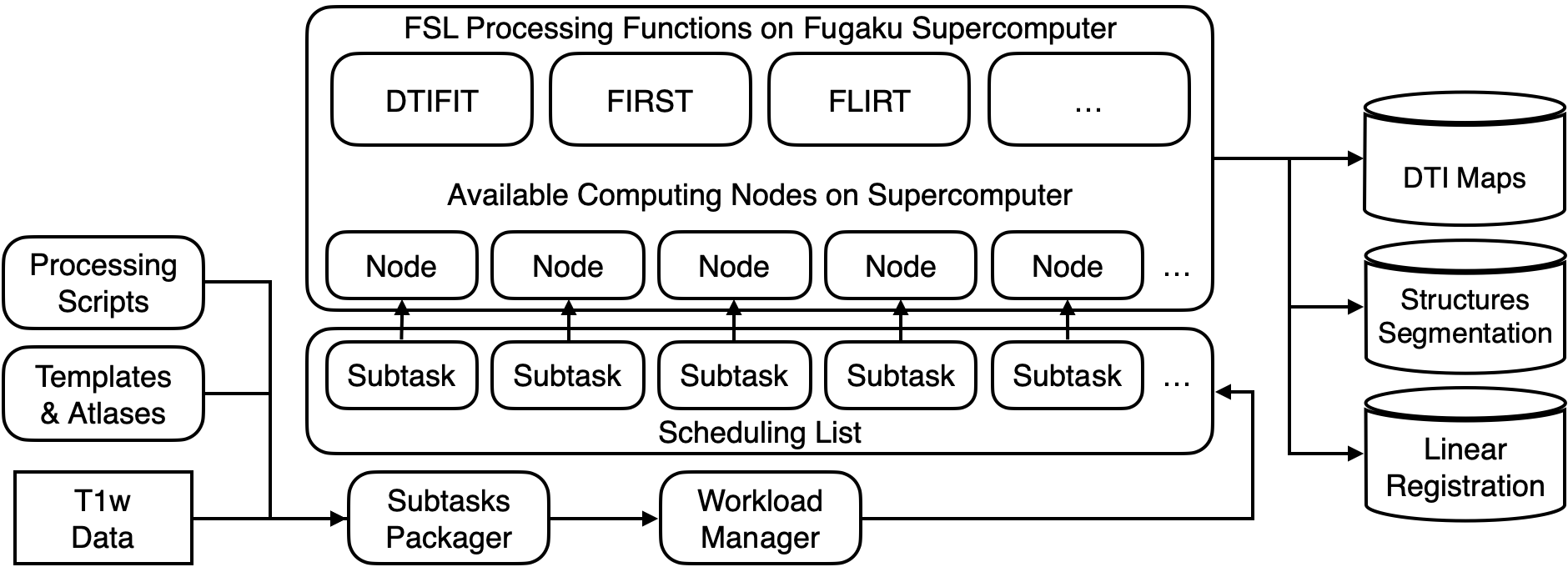}
    \caption{Data Parallelism Processing Framework}
    \label{fig:brain_atlas}
\end{figure}

\subsection{Verification of pipeline}
We verified the reproducibility of the pipeline implementation on the supercomputer Fugaku escribed in the previous section via comparing the results from the conventional node. First, to evaluate the reproducibility of voxel-wise DTI parameters, we assessed the correlation between DTI measures derived from the supercomputer and from the conventional node for each voxel. The $R^2$ was used as an index for evaluating the correlation. $R^2$ was defined as
$$R^2(y, \hat{y}) = 1 - \frac{\sum_{i=1}^{n} (y_i - \hat{y}i)^2}{\sum{i=1}^{n} (y_i - \bar{y})^2}$$
where $\bar{y} = \frac{1}{n} \sum_{i=1}^{n} y_i$ and $\sum_{i=1}^{n} (y_i - \hat{y}i)^2 = \sum{i=1}^{n} \epsilon_i^2$.\\\\
In addition, to compare the accuracy of DTI parameters estimation in local brain structures, we calculated the average DTI parameters for each deep gray matter structure defined in the diffusion-weighted imaging space and evaluated the correlation of the estimation results between general node and supercomputer as well as the voxel-wise approach.
\begin{figure}[h!] % image examples & compare
	\centering
    \includegraphics[width=0.8\textwidth]{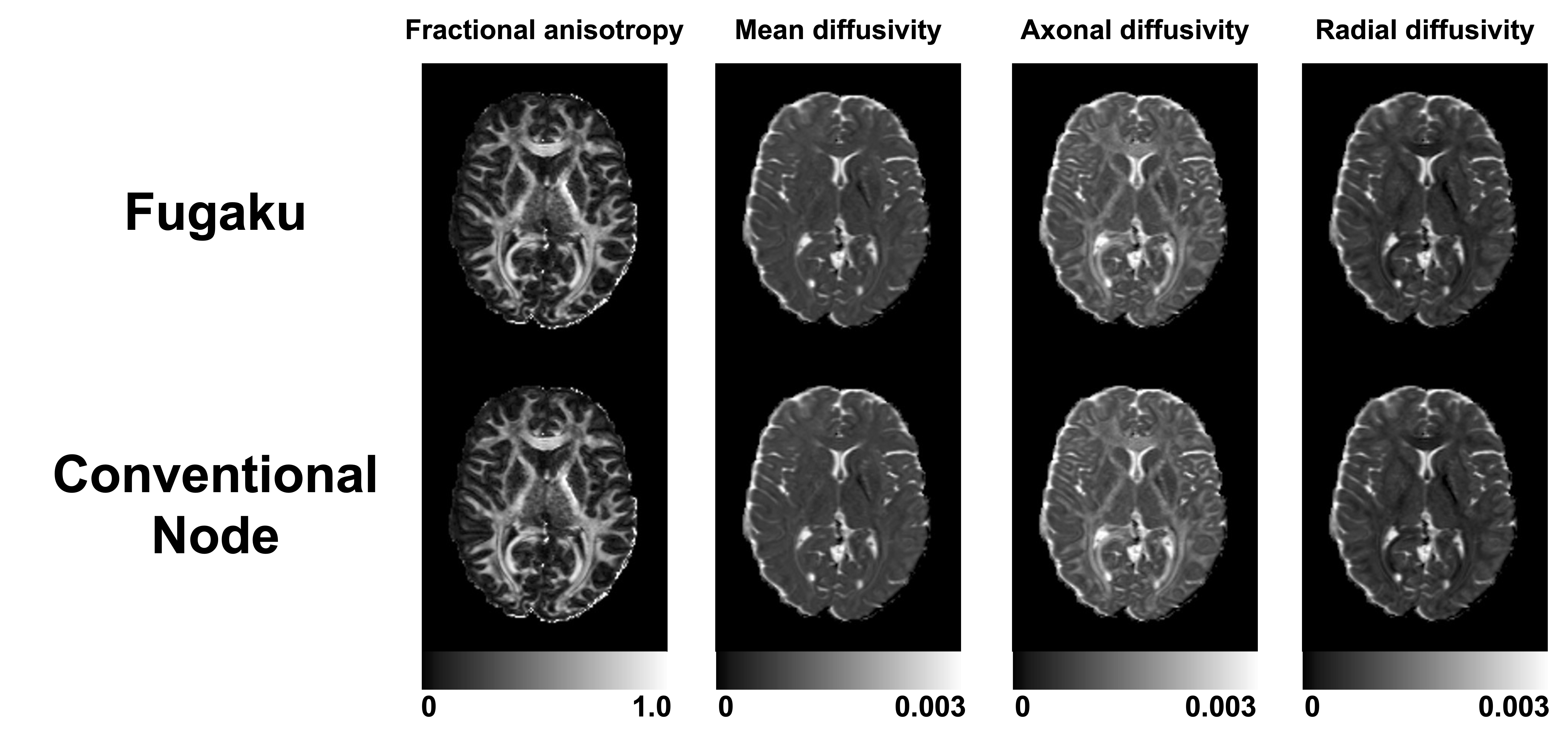}
    \caption{Comparison between Fugaku and Conventional Node}
    \label{fig:brain_atlas}
\end{figure}
Next, to confirm the segmentation accuracy of deep gray matter using FIRST, we used the dice index of ROIs defined by pair of two nodes.
The Dice similarity coefficient between two sets A and B is defined as:
$$\mathrm{dice}(A, B) = \frac{2 \times |A \cap B|}{|A| + |B|}$$
where $|A|$, $|B|$denotes the number of elements of node A and B, respectively, and $|A \cap B|$ represents the cardinality of the intersection of pair of two nodes. In simple terms, the Dice index is calculated as twice the number of elements in the intersection divided by the total number of elements in both sets.
\begin{figure}[h!] % image examples & compare
	\centering
    \begin{subfigure}{0.4\textwidth}
        \centering
        \includegraphics[width=1.0\textwidth]{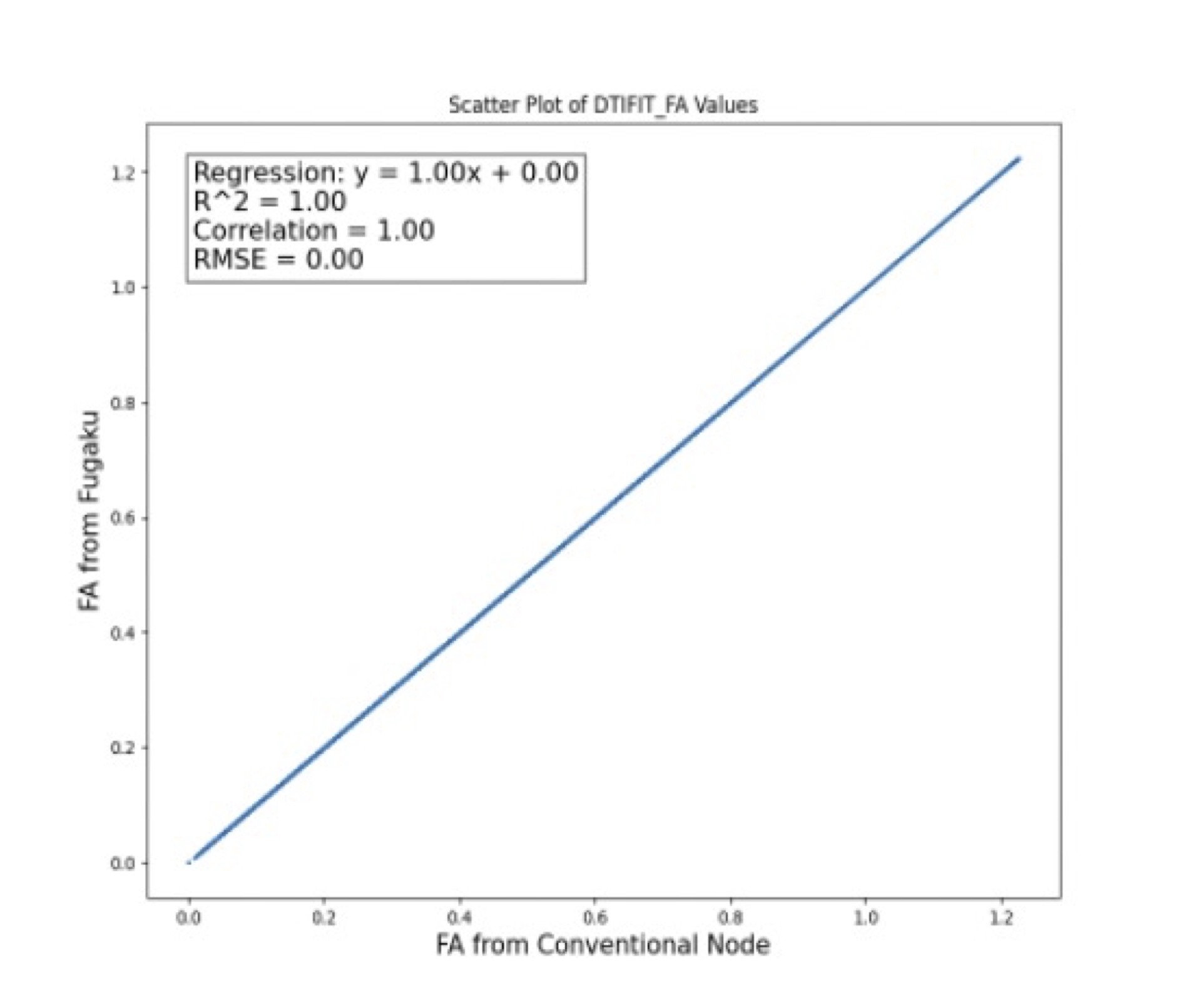}
        \caption{FA}
        \label{fig:FA}
    \end{subfigure}
    \begin{subfigure}{0.4\textwidth}
        \centering
        \includegraphics[width=1.0\textwidth]{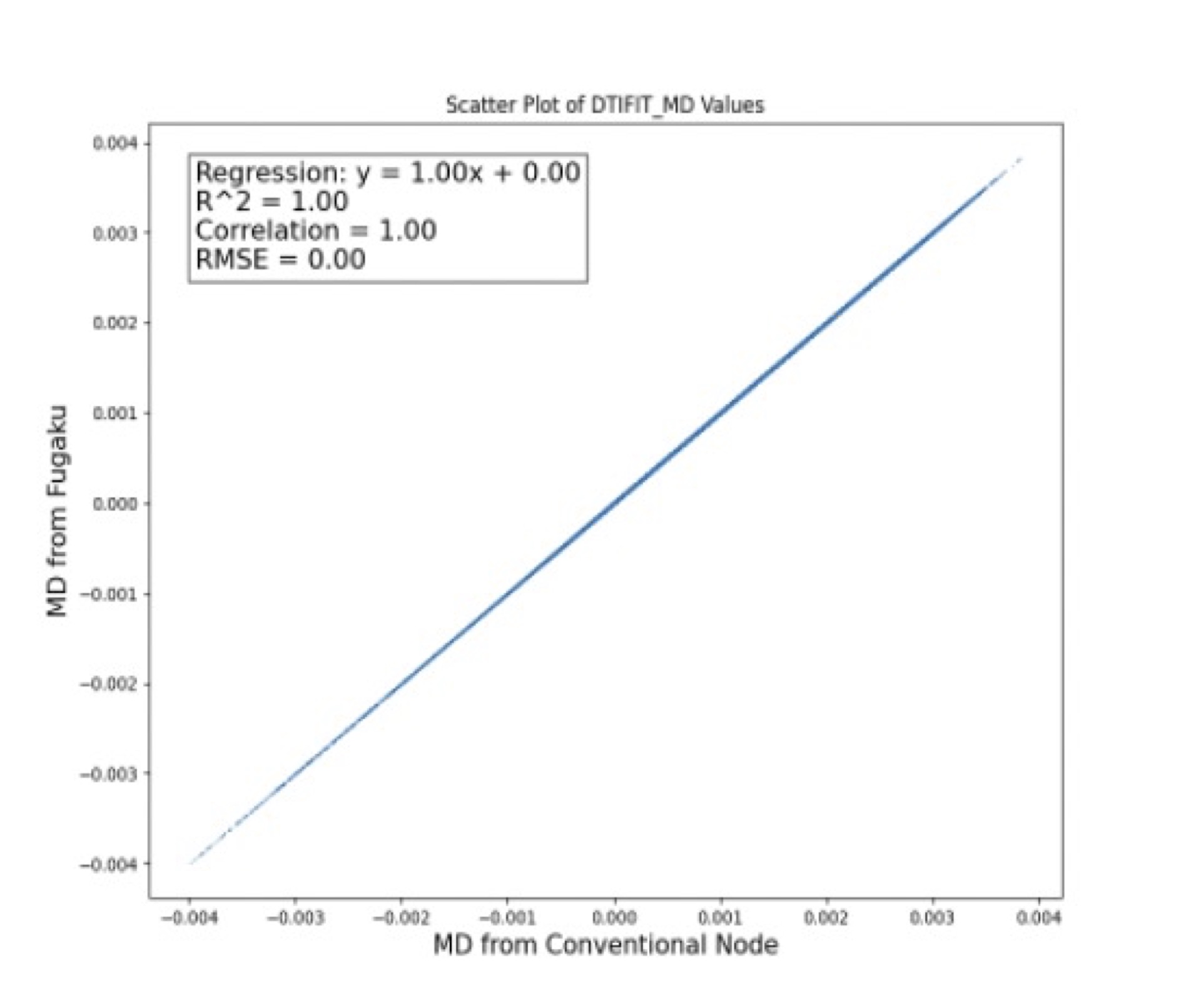}
        \caption{MD}
        \label{fig:MD}
    \end{subfigure}
    \begin{subfigure}{0.4\textwidth}
        \centering
        \includegraphics[width=1.0\textwidth]{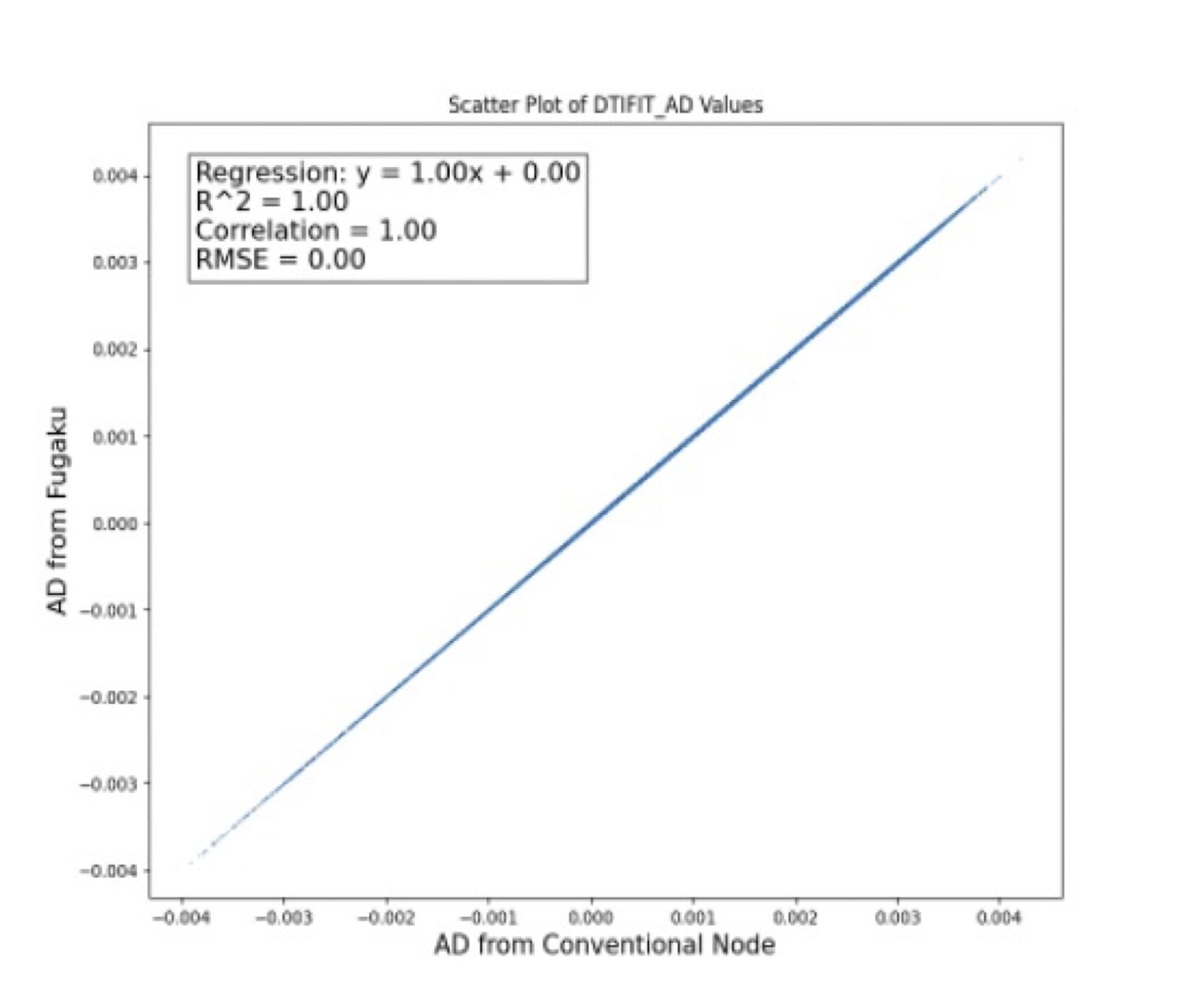}
        \caption{AD}
        \label{fig:AD}
    \end{subfigure}
    \begin{subfigure}{0.4\textwidth}
        \centering
        \includegraphics[width=1.0\textwidth]{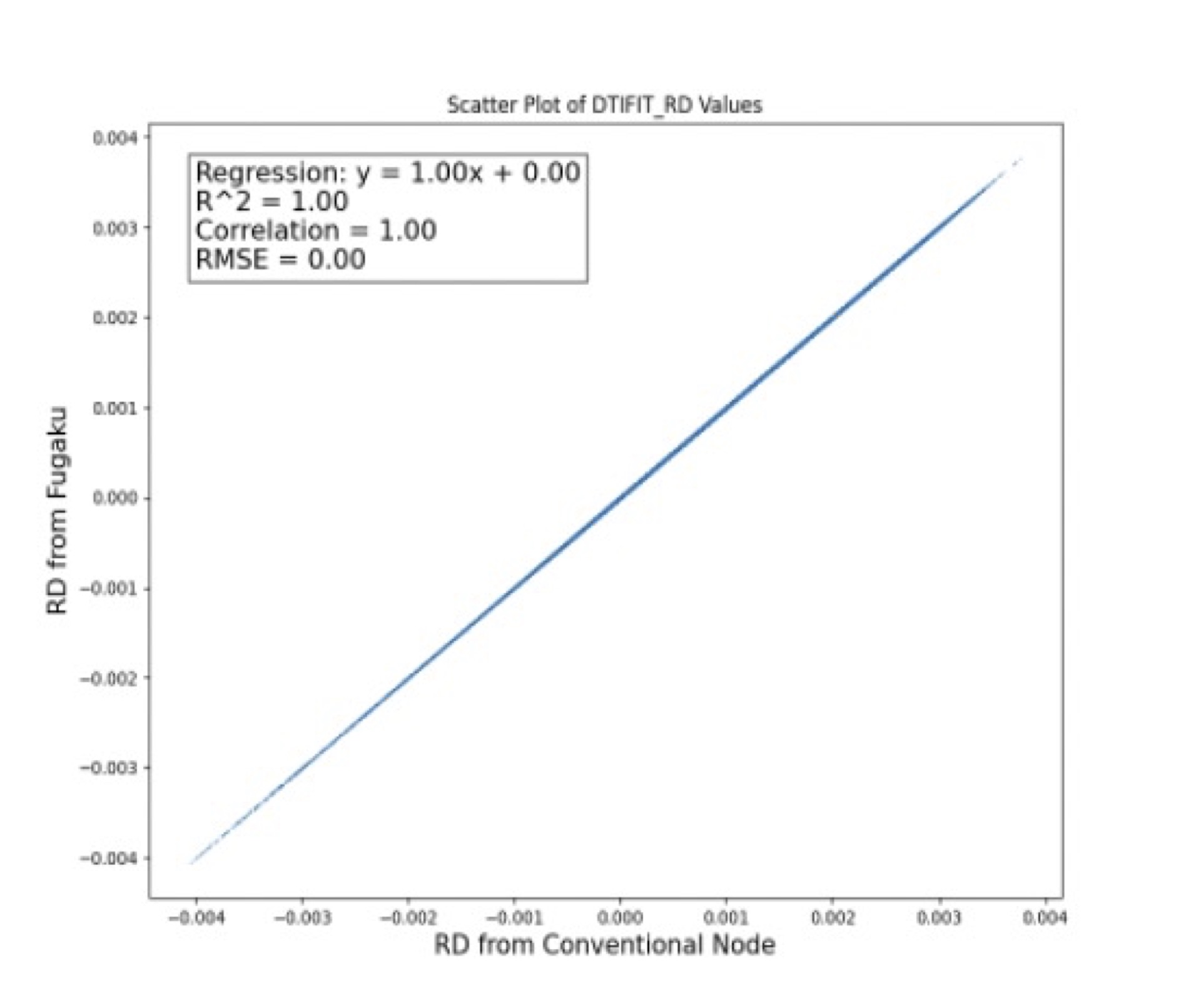}
        \caption{RD}
        \label{fig:RD}
    \end{subfigure}
    \caption{Scatter Plot of DTIFIT Values}
    \label{fig:dtifit_plot}
\end{figure}
\section{Results}
\subsection{Diffusion Tensor Imaging}
The voxel-wise correlation of DTI parameters between the supercomputer and conventional node was evaluated using the coefficient of determination ($R^2$). The scatter plots in Figure 2B show the voxel-wise correlation for each DTI parameter, with $R^2$ values of 1.0 for FA, 1.0 for MD, 1.0 for AD, and 1.0 for RD. These high $R^2$ values indicate an excellent agreement between the DTI results obtained from the supercomputer and the conventional node.
\begin{figure}[h!] % image examples & compare
	\centering
    \includegraphics[width=0.7\textwidth]{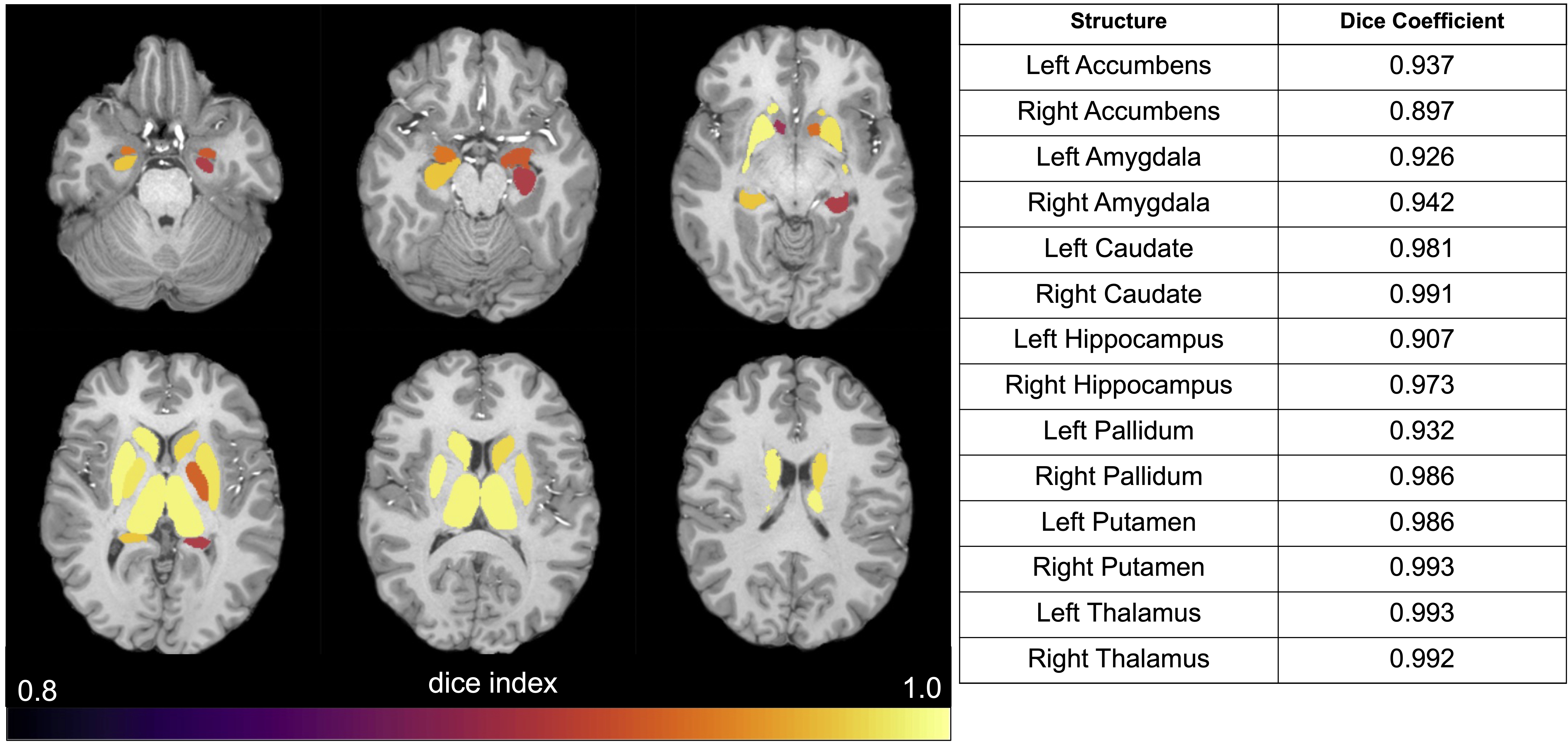}
    \caption{Structure and Dice Coefficient}
    \label{fig:dice}
\end{figure}
For the region-wise evaluation of DTI parameters in deep gray matter structures, scatter plots were created to assess the correlation between the supercomputer and conventional node results (Figure 4). The $R^2$ values were 1.0 for FA, 0.99 for MD, 0.99 for AD, and 0.99 for RD, demonstrating a high degree of consistency in the regional DTI parameter estimation between the conventional node and supercomputers including the Fugaku.
\begin{figure}[h!] % image examples & compare
	\centering
    \begin{subfigure}{0.4\textwidth}
        \centering
        \includegraphics[width=1.0\textwidth]{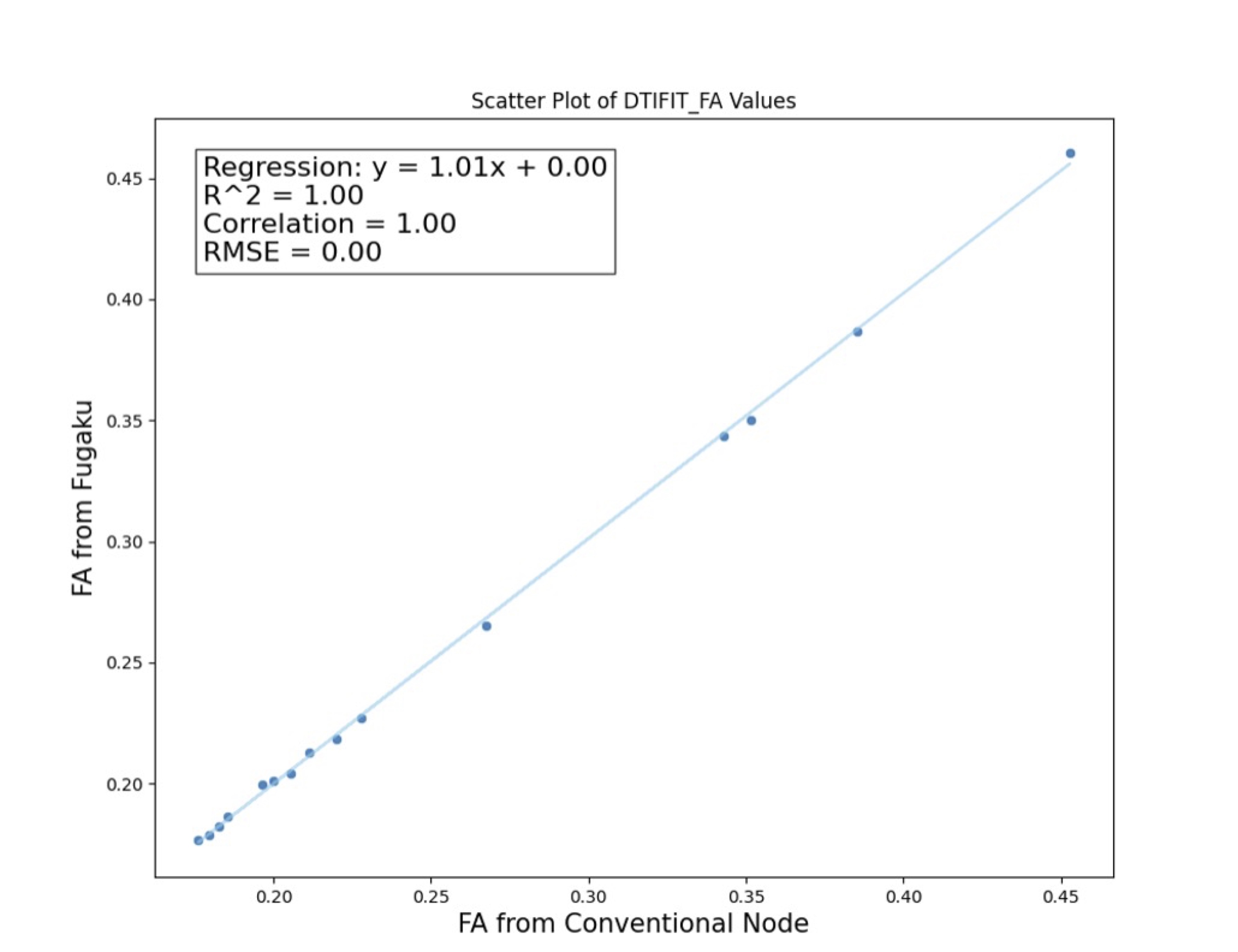}
        \caption{FA}
        \label{fig:dti_FA}
    \end{subfigure}
    \begin{subfigure}{0.4\textwidth}
        \centering
        \includegraphics[width=1.0\textwidth]{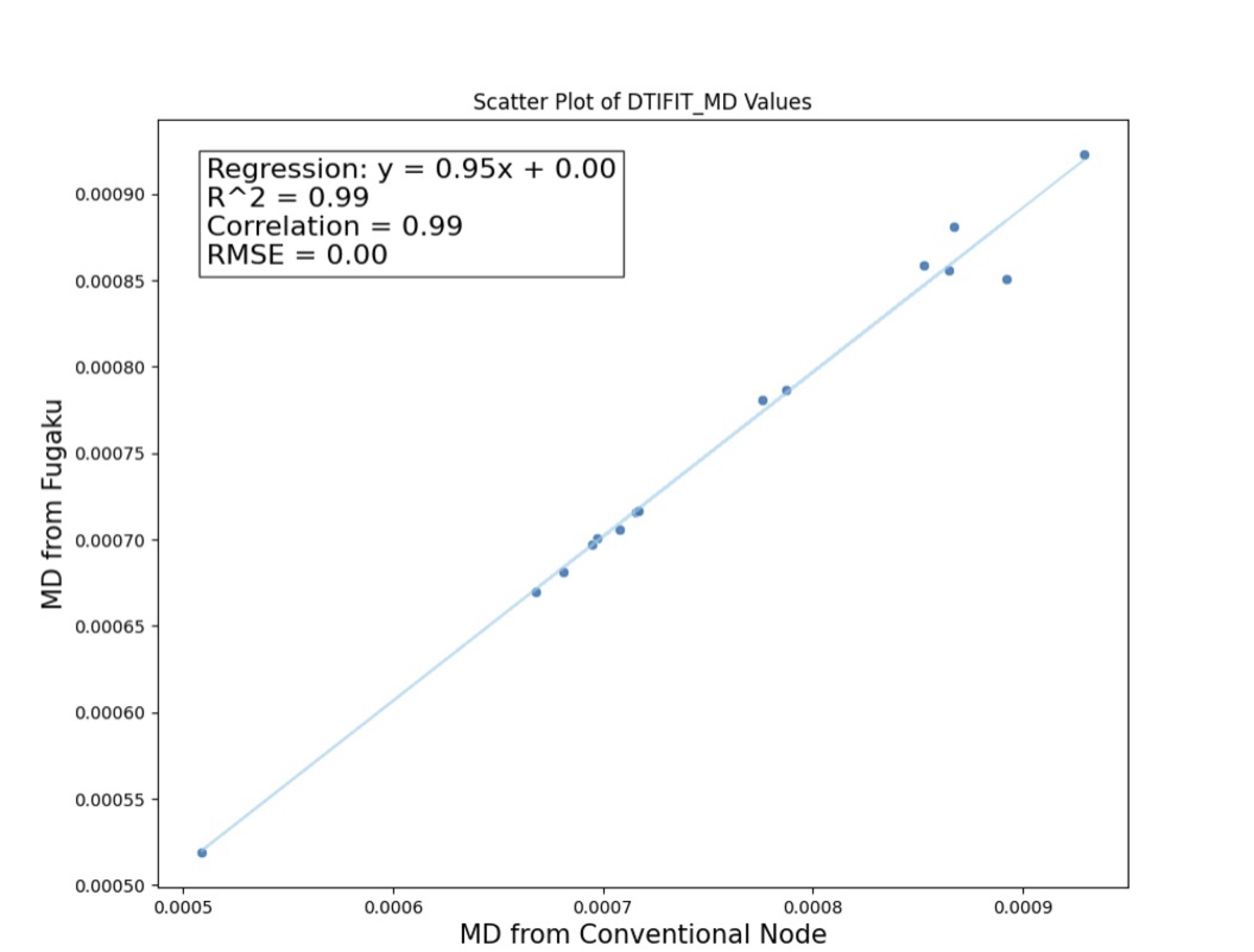}
        \caption{MD}
        \label{fig:dti_MD}
    \end{subfigure}
    \begin{subfigure}{0.4\textwidth}
        \centering
        \includegraphics[width=1.0\textwidth]{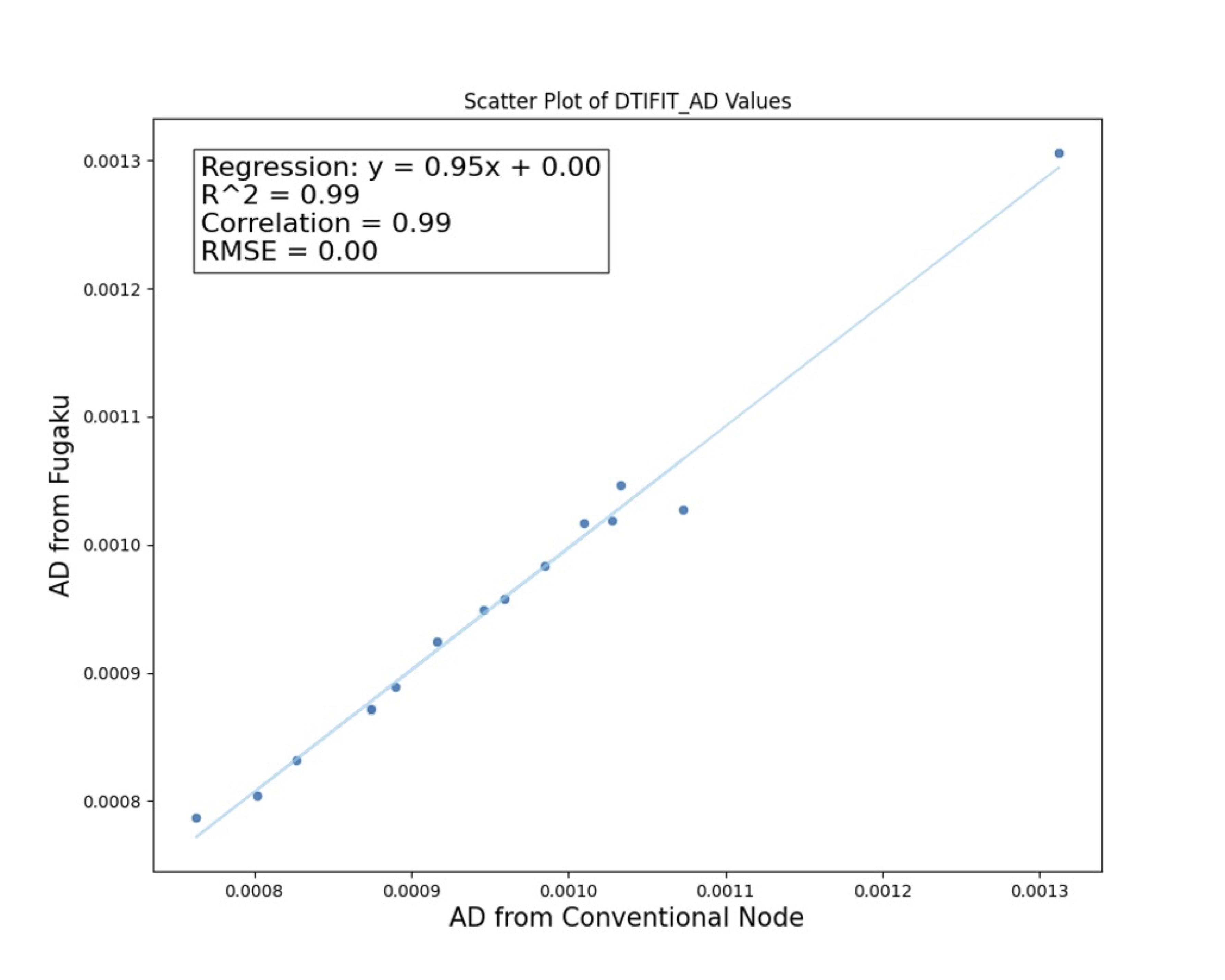}
        \caption{AD}
        \label{fig:dti_AD}
    \end{subfigure}
    \begin{subfigure}{0.4\textwidth}
        \centering
        \includegraphics[width=1.0\textwidth]{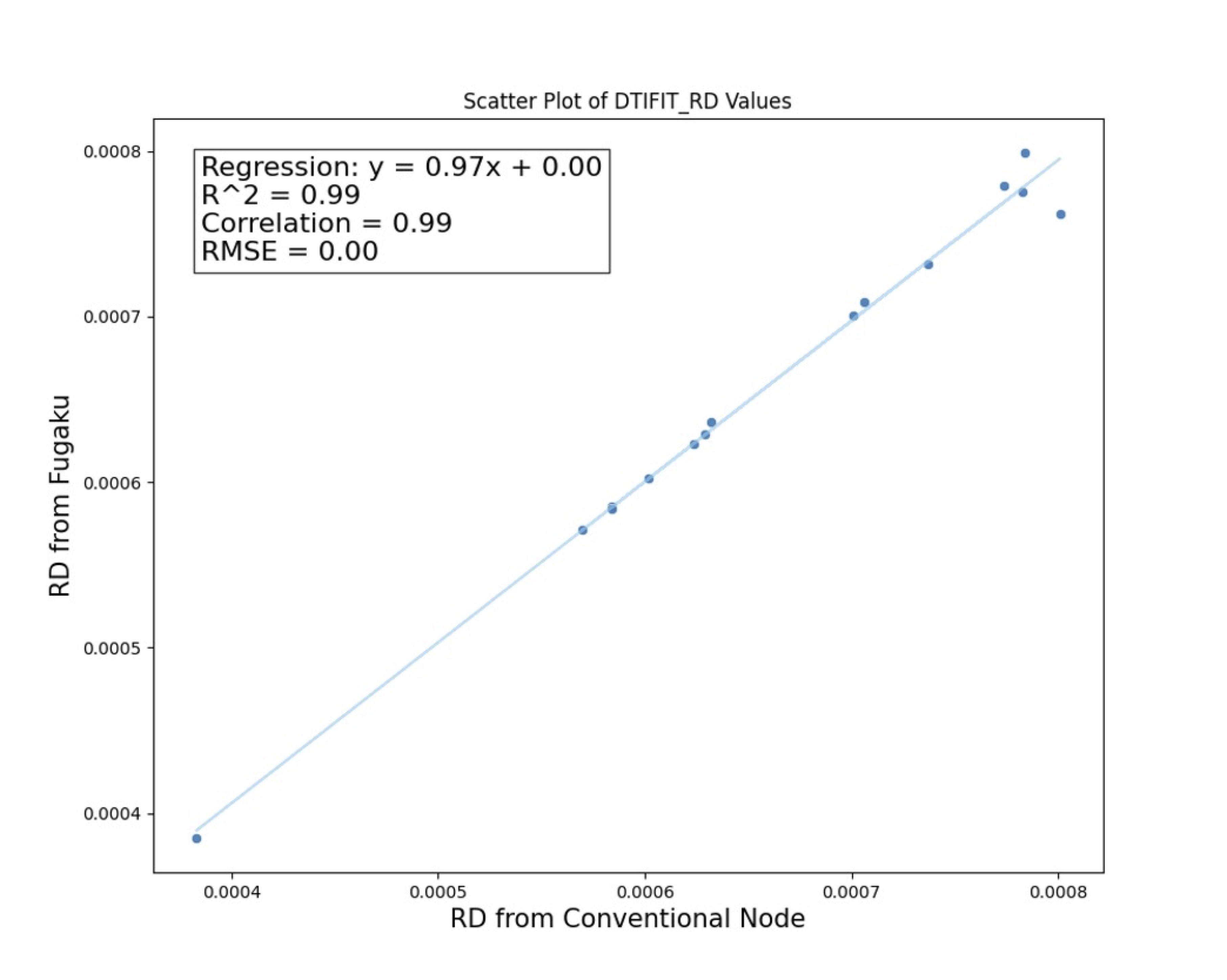}
        \caption{RD}
        \label{fig:dti_RD}
    \end{subfigure}
    \caption{Scatter Plot of DTIFIT Values}
    \label{fig:dti_plot}
\end{figure}
\subsection{Subcortical Sturctures Segmentation}
The deep gray matter segmentation results obtained using FIRST on the supercomputer were compared to those from the conventional node by calculating the dice index for each structure. As shown in Table 1, the dice indices ranged from 0.897 for the accumbents to 0.993 for the putamen and thalamus. These values indicate a high degree of overlap between the segmentation results from the two platforms.
\section{Discussion}
This study aimed to implement the FMRIB Software Library (FSL), a well-established brain diffusion MRI analysis pipeline, on the Fugaku supercomputer for the first time. The successful execution of FSL commands such as DTIFIT, FIRST, and FLIRT on brain MRI data of healthy young adults demonstrates the feasibility and potential of utilizing supercomputers for advanced brain image analysis.
\par
The implementation of FSL in the Fugaku supercomputer has increased the possibility of significantly reducing the time required to analyze and process large numbers of brain images based on large databases compared to conventional methods. This speed-up enables the analysis of larger sample sizes and more complex datasets in realistic timeframes, increasing the possibility of obtaining robust and reliable research results from large-scale studies based on multi-center collaborative databases, which cannot be achieved through small-scale, single-institution studies.
\par
Despite the different CPU architectures of Fugaku (ARM-based) and common commercially available workstations (Intel-based), the results obtained from both systems showed a nearly perfect correlation and high overlap in fractional anisotropy (FA), mean diffusivity (MD), axial diffusivity (AD), radial diffusivity (RD), and subcortical structure volumes. This finding strongly supports the reliability and validity of using supercomputers for brain image analysis.
\par
While some small-area variability was observed in the structural segmentation by FIRST, the degree of this variability was minimal\cite{VelascoAnnis2017}. Due to the limited sample size of this study, the potential reasons for these discrepancies and their impact on the overall results remain uncertain\cite{Smith2011}. Future studies with larger sample sizes are needed to investigate and address these variations.
\par
The successful implementation of FSL on Fugaku has significant implications for democratizing brain imaging research. By making large-scale neuroimaging studies more accessible and efficient, especially for smaller laboratories with limited computational resources, this advancement paves the way for researchers to conduct brain imaging studies based on large databases. This democratization of brain imaging analysis research will facilitate collaboration among researchers.
\par
Compared to previous attempts to use supercomputers for brain imaging analysis, the findings of this study are highly innovative, as we have developed a pipeline that can handle both T1-weighted images and diffusion MRI as input images. This advancement expands the scope of brain imaging research that can be conducted using supercomputers.
\par
The main limitation of this preliminary study is the single-subject design, which limits the generalizability of the results. Future validation in a larger number of subjects is necessary. Additionally, only the FSL pipeline was implemented in this study. Future attempts should be made to implement and validate other brain imaging analysis pipelines such as MRtrix3 and ANTs.
\par
In conclusion, this study successfully demonstrates the feasibility and potential of brain image analysis using FSL on the Fugaku supercomputer. The results showed high consistency with conventional processing methods and highlighted the significance of supercomputing in reducing processing time, enabling large-scale studies, and democratizing brain imaging research. Future research should focus on validating these findings in larger samples and integrating other widely used brain image processing pipelines into the Fugaku supercomputer.
\section{Acknowledgement}
This work was supported by JSPS KAKENHI Grant Grant-in-Aid for Scientific Research(B) Number 24K03005.
\bibliography{sample.bib} 
\bibliographystyle{plain}
\end{document}